\documentclass{aastex631}

\graphicspath{figures/}
\usepackage[margins]{trackchanges}
\usepackage{savesym}
\savesymbol{tablenum}
\usepackage{graphicx}
\usepackage{xcolor}
\usepackage{siunitx}
\usepackage{amsmath}
\usepackage{tabularx}
\usepackage{graphicx}
\usepackage[figuresright]{rotating}
\usepackage{color}  
\usepackage{subfiles} 

\newcommand{\au}{\ensuremath{{\textsc{au}}}}

\newcommand{\kms}{\ensuremath{{\rm\, km\,s^{-1}}}}

\newcommand{\forbid}[2]{[\textsc{#1}]\,#2\textrm{\AA}} 
\newcommand{\NeII}{\ensuremath{\textrm{Ne}\textsc{ii}}}

\newcommand{\VA}{V_{\textsc{a}}}
\newcommand{\VAb}{V_{\textsc{a}0}}
\newcommand{\VK}{V_{\textsc{k}0}}

\newcommand{\cs}{c_{\rm s}}
\usepackage{CJKutf8}
\newcommand{\given}{\,|\,}

\defcitealias{Nemer+2023}{NG23}
\defcitealias{BYGY16}{BYGY}
\defcitealias{Ercolano+Owen2016}{EO16}
\accepted{\today}

\begin{document}
\begin{CJK*}{UTF8}{gbsn}

\title{Constraining Protoplanetary Disk Winds from Forbidden Line Profiles with Simulation-based Inference}

\author[0000-0002-9220-0039]{Ahmad Nemer}
\affiliation{Center for Astrophysics and Space Science, New York University Abu Dhabi}

\author[0000-0003-1197-0902]{ChangHoon Hahn}
\affiliation{Department of Astrophysical Sciences, Princeton University, Princeton NJ 08544, USA}

\author[0000-0001-9592-4190]{Jiaxuan Li (李嘉轩)}
\affiliation{Department of Astrophysical Sciences, Princeton University, Princeton NJ 08544, USA}

\author[0000-0002-8873-5065]{Peter Melchior}
\affiliation{Department of Astrophysical Sciences, Princeton University, Princeton NJ 08544, USA}
\affiliation{Center for Statistics and Machine Learning, Princeton University, Princeton NJ 08544, USA}

\author[0000-0002-6710-7748]{Jeremy Goodman}
\affiliation{Department of Astrophysical Sciences, Princeton University, Princeton NJ 08544, USA}




\begin{abstract}

Protoplanetary disks are the sites of vigorous hydrodynamic processes, such as accretion and outflows, and ultimately establish the conditions for the formation of planets. The properties of disk outflows are often inferred through analysis of forbidden emission lines. These lines contain multiple overlapping components, tracing different emission regions with different processes that excite them: a high-velocity component (tracing a jet), a broad low-velocity component (tracing inner disk wind), and a narrow low-velocity component (tracing outer disk wind). 
They are also heavily contaminated by background spectral features. All of these challenges call into question the traditional approach of fitting Gaussian components to the line profiles, and cloud the physical interpretation of those components. 
We introduce a novel statistical technique to analyze emission lines in protoplanetary disks. Simulation-Based Inference is a computationally efficient machine learning technique that produces posterior distributions of the parameters (e.g. magnetic field, radiation sources, geometry) of a representative wind model when given a spectrum, without any prior assumption about line shapes (e.g. symmetry). In this pathfinder study, we demonstrate that this technique indeed accurately recovers the parameters from simulated spectra without noise and background. A following work will deal with the analysis of observed spectra.

\end{abstract}



\section{Introduction}\label{sec:intro}

Protoplanetary disks (PPDs) are composed of molecular gas and dust and encircle most low-mass pre-main-sequence stars, referred to as protostars. These disks play a pivotal role in the formation of planets, making them a subject of immense interest in research. Over time, PPDs disperse, typically within a few million years, likely due to a combination of stellar accretion, planet formation, and outflows \citep{Alexander+2014, Ercolano2017, Manara+2023}.
To investigate the characteristics and evolution of PPDs, researchers utilize tracers of disk winds that provide insights into various physical conditions and regions \citep{Gudel+2014}. Notably, emission lines from oxygen, sulfur, and neon have emerged as prime candidates for tracing disk winds, enabling the study of their physical properties through accurate modeling of line ratios. The detection of the $[\NeII]12.8\micron$ fine structure line has confirmed the existence of warm ionized disk winds, often exhibiting a blueshift relative to the stellar velocity, indicating an outflow moving toward the observer, while the emission from the receding outflow in the other half of the disk is obstructed by the dust \citep{Font+2004, Alexander2008, Gorti+2009, Pascucci+Sterzik2009}.\\

Detailed spectroscopic observations have revealed the presence of both fast outflows reaching speeds of hundreds \kms, as well as slower outflows with velocities $\lesssim10\kms$ \citep{Hartigan+1995}. In certain luminous Class I and II systems, the fast outflows develop into highly collimated jets that extend for parsecs and manifest as Herbig-Haro objects \citep[and references therein]{Bally2016,Pascucci+2023}. The rate at which mass is lost through these outflows is closely linked to the rate of accretion \citep{Hartigan+1995,Ellerbroek+2013,Rigliaco+2013,Natta+2014}. The acceleration of these jets is believed to involve magnetohydrodynamic (MHD) mechanisms, although the exact launch locations—whether in close proximity to the star or more broadly from the PPD—remain uncertain \citep{Frank+2014}. The lower-velocity component of the outflows is likely associated with the disk itself, supported by evidence such as inclination-dependent line widths, lower blue shifts, and higher estimated mass-outflow rates \citep{Fang+2018,Simon+2016,Banzatti2019}.\\

Despite the valuable information extracted from these observations, limitations arise from the scarcity of high-resolution infrared spectra, hindering a comprehensive understanding of PPDs, their evolutionary trends, and detailed line profiles. However, optical forbidden lines, including $\forbid{Oi}{6300}$ and $\forbid{Sii}{4068}$, offer an alternative avenue with abundant high-resolution optical spectra \citep{Hartigan+1995}. These lines exhibit distinguishable low-velocity components (LVCs), emission blueshifted by less than 30\kms, and high-velocity components~\citep[HVCs,][]{Hartigan+1995}. 
Recent studies have successfully modeled LVCs with magnetothermal wind models \citep{Nemer+2023}, although recent observations have shown that LVCs can be further deconstructed into broader components (BCs), with full width at half maximum (FWHM) of at least 40 \kms, and narrower components \citep[NCs,][]{Rigliaco+2013,Simon+2016,Fang+2023}. This decomposition into different components is usually done with a multi-Gaussian fitting, where the choice of the number and shapes of the Gaussians is chosen by minimizing the residual after fitting~\citep{Fang+2018}. \\

The exploration of the relationship between BC and NC FWHMs and the inclination of observed systems suggests that BC likely originates from a wind launched at radii of $0.05$-$0.5\au$ ($0.5$-$5\au$ for the NC), assuming Keplerian rotation. From these conclusions about emission regions, it was suggested that the NC traces an outer photoevaporative wind, while the BC has an MHD wind signature \citep{Simon+2016}. However, it is more likely that the emission originates from a range of radii spanning a range of temperatures and densities, and the observed components are clearly overlapped with no explicit separation into distinct components tracing distinct regions. Furthermore, the observed spectra are commonly contaminated with background spectral features that have to be processed before fitting the lines with any model which introduces further difficulty in analyzing the optical spectra \citep{Banzatti2019}. These issues put the Gaussian fitting procedure into question as it was suggested that analysis of the emission lines should be done in more robust way that is not based on the symmetry of the lines \citep{Ballabio+2020}. \\

In this paper, we introduce a Simulation-based Inference (SBI) approach to extract spectrum information on the basis of state-of-the-art wind models. With an innovative combination of convolutional neural networks and neural density estimation techniques we can interpret forbidden emission line spectra of PPDs, without the limitations of Gaussian profile fitting. We test the efficacy of our approach with simulated spectra in this paper and will employ the same methods to analyze a sample of observed spectra in a subsequent paper.

\section{Methods}\label{sec:methods}

We use the analytic wind models of \citet{Nemer+2023} to generate simulated spectra, then compress them with a neural spectrum encoder, and train a neural density estimator to carry out SBI as the mapping from model parameters to (compressed) spectra. Afterwards, we test the trained SBI network with a sample of simulated spectra (with known parameters) that it has not seen during training. We summarize the different steps below. 

\subsection{The wind model}\label{subsec:windmodel}

In accordance with the methodology outlined by \cite[hereafter BYGY]{BYGY16}, we undertake the solution of the steady-state, axisymmetric ideal magnetohydrodynamic (MHD) equations governing fluid flow along each poloidal magnetic field line, under the condition of a specified density $\rho_0$ and poloidal Alfv\'en speed $\VAb=B_{\textsc{p}0}$ at the base of the wind. These winds possess a finite sound speed $\cs$, which remains spatially invariant along each magnetic field line with radial variation across lines. Termed ``magnetothermal'', these winds are characterized by their acceleration resulting from a confluence of magnetic forces (represented by $\VA$) and gas pressure (represented by $\cs$).
The configuration of the poloidal magnetic field is delineated by straight field lines that intersect the surface of the accretion disk at an inclination angle $\theta'$ from the disk midplane, with the specific intersection points denoted by cylindrical coordinates $(R_0,z_0)$. Anticipating these intersections to lie several scale heights above the midplane, where FUV radiation from the star can penetrate and ionize the gas sufficiently to couple it with the field, we set $z_0=0.15 R_0$ to approximate this condition.
The dynamics of the flow along each magnetic field line are governed by four dimensionless input parameters $(\theta',z_0/R_0,q,\VAb/\VK,\cs/\VK)$ and three dimensional scaling factors $(R_0,\rho_0,\VK)$, with the latter representing the Keplerian orbital velocity at the base of the wind, $\VK=\sqrt{GM_*/R_0}$. Notably, for $q<1$ the scaling of the poloidal magnetic field strength with radius smoothly transitions from $B\propto R^{-1}$ to $B\propto R^{-2}$ at $R\gtrsim R_0/q$.

\subsection{Postprocessing with Cloudy}

We utilize the radiative transfer software Cloudy \citep{Ferland+2017} to analyze the density and velocity distributions within our models of stellar winds. This analysis, coupled with appropriate energetic photon luminosities, allows us to compute the emission of the \forbid{Oi}{6300} line, following the methodology outlined in \citet{Nemer+2023}. Our focus lies predominantly on the inner-to-intermediate regions of the disk, spanning distances from (0.04-4 \au{}), as these regions contribute the majority of the emission observed in the line. The computations are conducted on a spherical grid, with the radial cell size being variable and automatically adjusted by Cloudy to optimize the accuracy of the calculations. The grid comprises 361 points uniformly distributed in the $\sin\theta\in[0,1]$ coordinate space within the range [0,1]. Under the assumption of the disk midplane being optically thick, we exclude the lower half of the disk from our analysis. To determine the luminosities of the emission lines, we integrate their emissivities along radial paths. Each such path is associated with its unique density profile and is integrated over one hemisphere $(0\le\theta\le\pi/2)$, a condition pertinent to optically thin emission lines.

We construct a three-dimensional axisymmetric volume and conduct computations to derive the spectral profiles of our lines under various lines of sight, corresponding to different inclinations. The resultant line profile is intimately linked to the local gas temperature and flow velocity within the volume. The emitted radiation from each cell undergoes broadening based on the local gas temperature, and its distribution across velocity bins is determined by the projections of the flow onto the observer's line of sight at a specified inclination angle. Aggregating the contributions from all velocity bins yields the composite line profile, which is subsequently adjusted to conform to the calculated emission luminosity. To address instrumental effects, we incorporate convolution with a Gaussian function characterized by a standard deviation of $\sigma=8\kms$, mirroring the resolution capabilities of observational instruments.

In total, we generate 27,860 simulated \forbid{Oi}{6300} spectra from sampling 9 input parameters (as described in the following section and shown in Fig. \ref{fig:fig1}): the Alfv\'en velocity $v_A$, the sound speed $c_s$, the geometry of the poloidal field lines $\theta_B$, the inner radius of the disk $R_{\rm in}$, the luminosity of various radiation sources $L_{\rm stell}, L_{\rm FUV}, L_{\rm EUV}, L_{\rm Xray}$, and inclination of the system $inc$. 

\subsection{Simulation-Based Inference using Normalizing Flows}
Our goal is to infer the posterior distribution of the 9 model parameters (PPD attributes),
$\theta=\{v_A, c_s, \theta_B, R_{\rm in}, L_{\rm stell}, L_{\rm FUV}, L_{\rm EUV}, L_{\rm Xray}, inc\}$, given a spectrum of forbidden emission lines, $x$: 
$p(\theta\given x)$. 
With the conventional Bayesian probabilistic inference approach, the posterior would be evaluated using
Bayes' rule: $p(\theta\given x) \propto p(\theta) \,p(x\given\theta)$. 
$p(\theta)$ is the prior distribution and 
$p(x\given\theta)$ is the likelihood, which is often
evaluated assuming a Gaussian functional form. 
Then, the posterior is inferred by sampling it using Markov Chain Monte Carlo (MCMC) methods. 
Sampling a 9-dimensional posterior distribution with MCMC 
requires {\em many} ($\gg 10,000$) evaluations of the posterior.
Each evaluation in our case involves running a full forward model that requires 400-500 CPU hours. 
This makes the conventional inference approach computationally infeasible.

SBI offers an alternative approach. 
The latest SBI methods use neural density estimators (NDEs)
to directly approximate the posterior using a training 
dataset, $\{(x_i, \theta_i)\}$~\citep[e.g.,][]{alsing2019, wong2020, dax2021, zhang2021, Hahn2022_simbig}. 
With NDEs, we can efficiently and accurately estimate high-dimensional posteriors with just a few thousands of training data~\citep[e.g.,][]{Hahn2022_simbig}.
This approach has additional benefits.
SBI does not make any assumptions about the functional form of the likelihood.
Furthermore, once the neural posterior estimator is 
trained, generating the posterior for a new spectrum requires no additional model evaluations and training.

Below we describe our SBI approach in detail. 
We follow the SBI approach in \cite{Hahn2022_sedflow} with an added data compression/feature extraction step that reduces the dimensionality of the spectrum and improves the overall performance of the SBI model.

\subsubsection{Prior Distribution}
\begin{figure}[ht]
\centering
\includegraphics[width=0.8\textwidth]{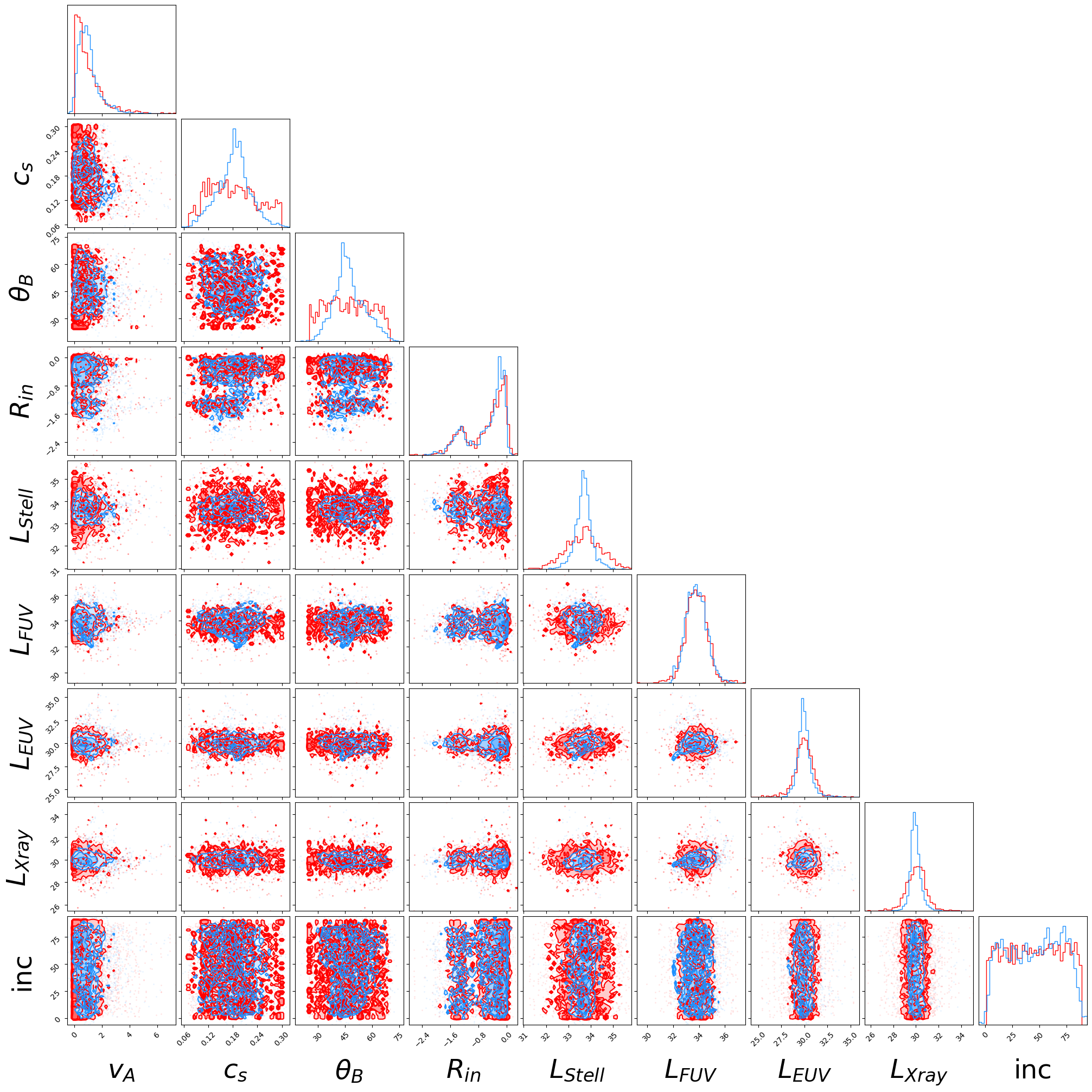}
\caption{The distribution of model parameters, 
$\theta=\{v_A, c_s, \theta_B, R_{\rm in}, L_{\rm stell}, L_{\rm FUV}, L_{\rm EUV}, L_{\rm Xray}, inc\}$. The blue contours show the distribution of physical parameters used to generate the training spectra. The red contours instead show the latent space after compressing the spectra with our trained encoder {\sc Spender}. The agreement between the two distributions illustrates that the encoder successfully extracts the essential information about the parameters from the full spectrum.}
\label{fig:fig1}
\end{figure}

In SBI, the prior distribution is set by the distribution of the parameters in our training data.
In our case, the prior distribution is not uniform as there are physical limitations that prohibit certain combinations of input parameters. The choice of the prior is based on the exploratory work by \citet{Nemer+2023} done with this wind model: 

\begin{itemize}
\item For the radiation sources, $L_i$, the central value of the prior distribution was found to reproduce the average \forbid{Oi}{6300} observed line luminosity, ratio to \forbid{Oi}{5577}, and the shape of the LVC. The width of the prior distribution for these sources is chosen to mimic the spread in the quantities as reported by observations \citep{Rigliaco+2013}. 
\item The sound speed, $c_s$,  chosen to reflect a temperature $\sim5,000 - 10,000K$ in the line emitting region (\citetalias{Ercolano+Owen2016}) since the temperature is 
prescribed to be constant along the field lines in the wind solution. 

\item The inclination, $inc$, is chosen to be uniform since there is no specific preferred inclination for PPDs. 
\item The inner radius of the model, $R_{\rm in}$, is chosen to be bimodal to reflect both classical PPD systems, with a primordial disk, and disks with inner cavities where there is little to no emission from inner radii. 

\item Disk magnetic field, on the other hand, has no solid measurements, so the magnitude and the geometry are somewhat ambiguous. 
Hence, $v_A$, which is a proxy for the field magnitude, is chosen within a range that would produce reasonable line width in accordance with observations; we did sample outlier field values that would produce extreme line widths as this has also been seen in observations \citep{Banzatti2019}. 
\item Poloidal field lines parallel, or close to parallel, to the rotation axis or the midplane can only accelerate the gas in extended regions ($>5\au$) which is outside the expected emission region \citep{Fang+2023}. So, the choice of field angles $\theta_B$ is constrained between $20^o$ and $70^o$
\end{itemize}
We present the distribution of parameters in the training data, i.e., our prior distribution (blue), in Fig.~\ref{fig:fig1}. 

\subsubsection{Data Compression}

The simulated spectrum has 4,000 spectral elements centered at the central wavelength. However, the success of past works in decomposing the spectra into Gaussian components strongly suggest that it can be efficiently 
expressed in a much lower dimensional space, without the loss of significant information.
In this work, rather than using Gaussian decomposition, we use the  
convolutional encoder from {\sc Spender}~\citep{Melchoir+2023,Liang2023a,Liang2023b}, an
autoencoder neural network architecture designed to extract compact 
representation of spectra. 
With the {\sc Spender} encoder, we can efficiently extract relevant features
of the spectra while also relaxing any assumptions on the shape of the 
different components.

{\sc Spender} employs a convolutional encoder $g(x)$ with an attention layer 
to identify the most prominent patterns or ``features'' in the spectrum $x$.
The input spectrum of the encoder first goes through
three convolutional layers, which account for correlations among the dominant features and segment the spectrum in wavelength. 
The attention mechanism then identifies the most influential spectral features, which are finally passed through a fully 
connected multilayer perceptron to produce the 
compressed spectrum in latent parameter space. In \cite{Melchoir+2023}, the compressed spectrum is passed through a decoder and the {\sc Spender} autoencoder is trained by minimizing the reconstruction loss between the input spectrum and the output of the decoder.

In this work, we only use the {\sc Spender} encoder without invoking the decoder.  
We train it to predict the input model parameter values used to generate the simulated spectrum. 
More specifically, we minimize the L2 norm between the input and predicted model parameters. As a result, our compression reduces the spectrum from 4,000 spectral elements to $N=9$ dimensions, equivalent to the number of input parameters. In other words, the encoder acts as a data compression and initial parameter regression step. By training our encoder to predict the model parameters, we design the encoder to extract the features with the most constraining power on the model parameters. A similar data compression approach was used in \cite{Chen2023,lemos2023}.

To train the encoder, we first reserve 15\% of the 
data for testing. 
Then, we split the rest of the data into an 80\% training 
and 20\% validation sets.
We train the encoder for 100 epochs with the Adam optimizer \citep{Adam}, the 1Cycle schedule \citep{Smith2017} with a maximum learning rate of $5\times10^{-3}$, and a batch size of 128. On an NVIDIA V100 GPU, the training takes 5 minutes approximately.
Our final trained encoder has comparable training and validation losses. 
This suggests that the network is not overfitting and  
we conclude that the training procedure is overall stable.
As shown in Fig.~\ref{fig:fig1}, the predicted distribution of latent parameter/compressed spectrum for the test sample (red) is similar to the input parameters of the training sample (blue).
This demonstrates that the encoder successfully compresses
the spectrum into a lower dimensional latent space 
that encodes the important features in the spectrum. 

\subsubsection{Simulation Based Inference}

From the compressed spectrum $s=g(x)$, we obtained the parameter posteriors $p(\theta\given s)$ using SBI based on ``normalizing flows'' \citep[e.g.,][]{Tabak2010,Tabak2013,Kobyzev2019}. In short, normalizing flow models utilize an invertible bijective transformation $f$ to map a complex target distribution to a simpler base distribution. This transformation needs to be invertible and have a tractable Jacobian for evaluation of the target distribution from the base distribution. A neural network, with parameters $\phi$, is trained to represent $f$.
Among various normalizing flow models~\citep[e.g.,][]{Germain2015, Durkan2019}, we use Masked Autoregressive Flow~\citep[MAF;][]{Papamakarios2017} models implemented in Python package \texttt{sbi}\footnote{\url{https://sbi-dev.github.io/sbi/}} \citep{tejero-cantero2020sbi}, similar to \cite{Hahn2022_sedflow}.

Our goal is to train a MAF model, $q$, with hyperparameters $\phi$, that accurately estimates the posterior, i.e. the conditional probability distribution of the parameters given the (compressed) data: $p(\theta\given s) \approx q_\phi(\theta\given s)$. 
We do this by minimizing the KL divergence between 
$p(\theta\given s)$ and  $q_\phi(\theta\given s)$, 
which is equivalent to maximizing the total log
likelihood, $\sum_i \log q_\phi(\theta_i\given s_i)$ over the training set.
In practice, we first split the training data into a training and validation set with a 90/10 split.
Then we use the Adam optimizer with a learning rate of $5 \times 10^{-4}$ to maximize the log likelihood. 
We use early stopping to prevent overfitting: we evaluate the total log likelihood on the validation data at every training epoch and stop the training when the validation log likelihood fails to increase after 20 epochs.
We determine the architecture of our MAF (e.g., number of blocks, hidden layers) through experimentation and selecting the model  that produces the highest validation log likelihood.
Our final MAF model has 3  Masked Autoencoder for Distribution Estimation~\citep[MADE;][]{Germain2015}
blocks, each with 2 hidden layers and 80 hidden units. 

To apply SBI directly to observations, we require simulated spectra with realistic noise and observational systematics. Available PPD observations are usually processed for background spectral features (including photospheric and telluric absorption lines) utilizing stellar templates of known systems and precise measurements of the system's radial velocity, as outlined by \citet{Banzatti2019,Fang+2018}. 
At present, however, we lack sufficient information about these processing steps and their uncertainties to include their effects on our mock spectra. Consequently, in this paper, we do not add noise and background to the simulated spectra and focus on demonstrating the 
overall feasibility of our SBI framework for analyzing forbidden line profiles. In the subsequent paper, when we apply our framework to observations, we will supplement our model with a realistic background
spectral features model that is tailored to the specific observations.

\section{Results}\label{section:results}

\begin{figure}[t]
\centering
\includegraphics[width=\textwidth]{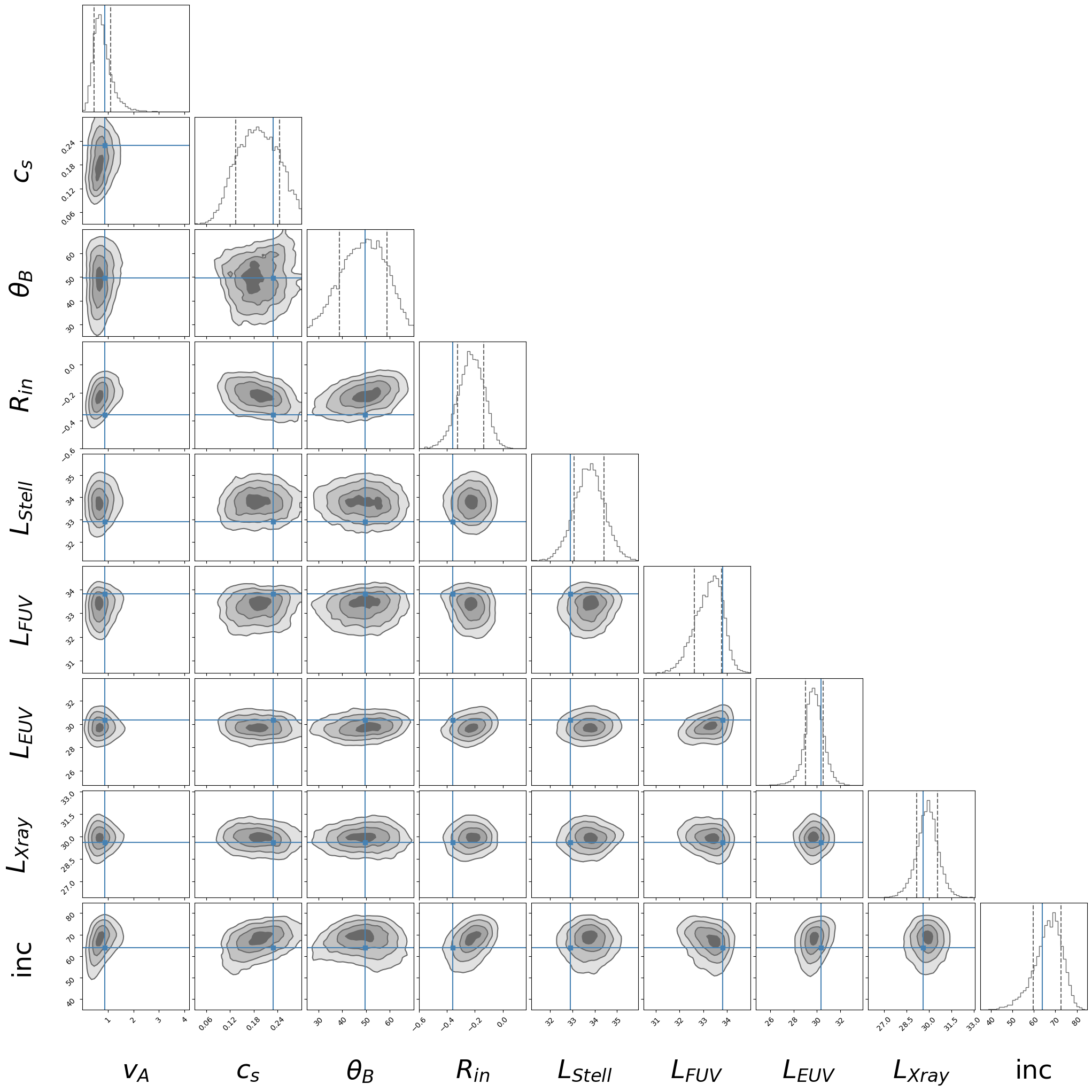}
\caption{The posterior estimate $q_\phi (\theta\given x)$ for a randomly selected test spectrum. The inferred posterior using SBI is consistent with the truth parameter values (blue).}
\label{fig:fig2}
\end{figure}

To demonstrate how the posterior of a spectrum is estimated using SBI, we show in Fig. \ref{fig:fig2} the posterior estimate $q_\phi(\theta\given x_i)$ for a randomly selected test spectrum. We overlay the true input parameters of the test spectrum (vertical and horizontal blue lines) for reference. The estimated $q_{\phi}(\theta\given x)$ is in excellent agreement with the truth. The spread in the distribution is a consequence of the loss of information, either in the mapping from parameters to spectra or in the encoding from spectra to initial parameters. \textbf{To our knowledge, this is the first time a method is presented to robustly connect wind model parameters to observed line shapes without the need for parametric fitting with Gaussians.} Each off-diagonal panel presents the marginalized 2D posterior of different parameter pairs, where the contours represent the 68, 95, and 99 percentiles. The diagonal panels present the marginalized 1D posterior of each parameter with the 16 and 84 percentiles (dashed) marked.

\begin{figure}[t]
\centering
\hspace*{-0.75in}
\includegraphics[width=\textwidth]{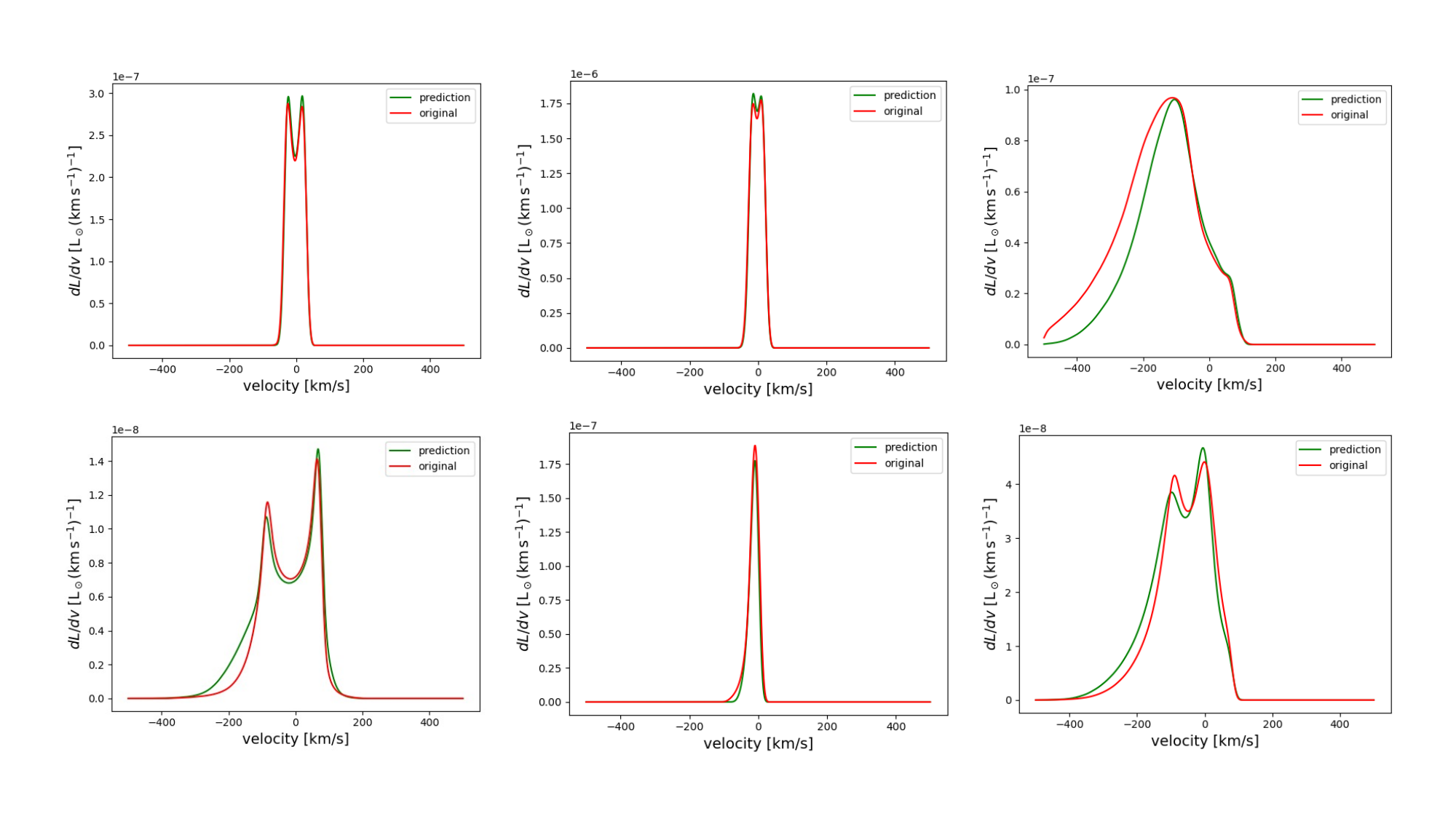}
\caption{Comparison between the simulated spectrum (red) and the forward-modeled spectrum of the MAP estimate from $q_\phi$ (green). They are in overall good agreement with each other. 
}
\label{fig:fig3}
\end{figure}

To further test whether $q_{\phi}(\theta\given x)$ robustly approximates the true posterior $p(\theta\given x)$, we take the 
maximum a posteriori (MAP) of the estimated posterior and perform forward modeling on it to predict the corresponding spectrum. We then examine how closely the MAP simulated spectrum reproduces the original spectrum for a given sample. 
This suggests that the parameters, in which SBI places the highest confidence, adequately represent the original spectrum.
To estimate the MAP, we rank the posterior samples by their log probability and select the parameters with the highest value. 
In Fig.~\ref{fig:fig3}, we present the comparison between the line profiles of the simulated spectra (red) and the MAP spectra (green). 
We find good agreement between the spectra despite their complex, non-Gaussian shape. It is noteworthy that three of the modelled spectra in Fig.~\ref{fig:fig3} show a slight double peak, while one shows an extreme double-peaked profile; a similar feature was reported in previous modelling work \citep{Weber+2020}. The double-peaked profile is due to the rotational motion of the gas, and is expected from any system that exhibits Keplerian motion, but is rarely observed in PPDs. The slightly-double-peaked profiles are expected to be smoothed out when noise and background features are added. Nevertheless, even if the predicted double-peaked emission does not represent any real situation, it helps to have these out-of-sample predictions to avoid overfitting the neural networks with biases from our prior knowledge of the observations.

\begin{figure}[t]
\centering
\includegraphics[width=0.7\textwidth]{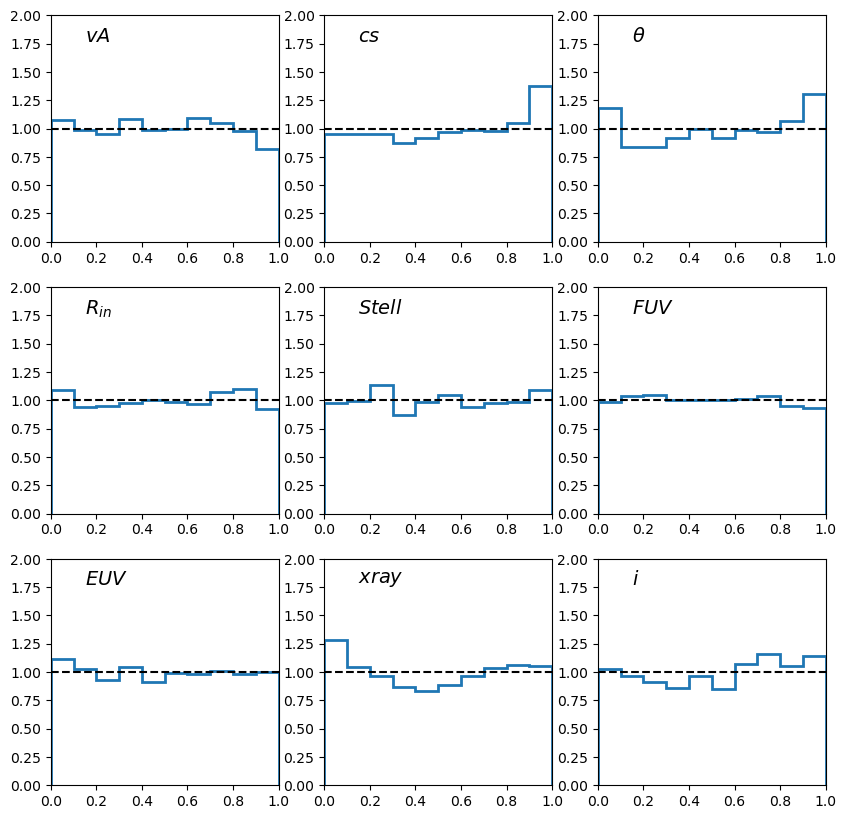}
\caption{SBC plot of the model posteriors for 4179 simulated spectra. The histogram in each panel represents the distribution of the rank statistic of the true value within the marginalized model posterior (blue) for each parameter. For the true posterior, the rank statistics will have a uniform distribution (black dashed). The rank statistic distribution of the model is nearly uniform for all of the parameters. }
\label{fig:fig4}
\end{figure}

Next, we validate the accuracy of $q_\phi$ using simulation-based calibration \citep[SBC,][]{Talts2018}. SBC examines the distribution of the rank statistics of the true parameter values within the marginalized posteriors. In Fig. \ref{fig:fig4}, we present the SBC of each model parameter for the normalizing flow posteriors (blue) using 4179 test samples. The shape of this distribution can tell us about the quality of predicted posteriors compared with the true ones, including whether it is biased and whether it is over- or under-confident. Any asymmetry in the distribution implies that the posterior estimates are biased. For example, a U-shaped distribution where the true parameter values are more often at the lowest and highest ranks means that the posterior estimates are narrower than the true posteriors. 
We do not find these features for any of the model parameters in Fig. \ref{fig:fig4}. 
Hence, SBI provides unbiased and accurate estimates of the true posteriors (black dashed) for all model parameters. \\

\section{Summary and Discussion}\label{section:discussion}

The motivation behind the analysis done in this paper is to provide a more capable framework for analyzing optical observations of PPDs. The line profiles and ratios provide key information about the emitting region which helps understand the physical properties of the system at different evolutionary stages. Conventionally, observations of optical forbidden emission lines were analyzed using Gaussian fitting \citep{Fang+2018,Banzatti2019}. Notably, observations exhibit multiple components that are assumed to trace distinct emission regions.

Current models of PPDs indeed confirm the existence of multiple emission regions, and as a result, multiple components \citep{Weber+2020}. However, our models show that the lines are typically asymmetric, broad, and trace a range of physical properties (temperature, density, radiation). Moreover, the observed spectra are usually contaminated with background spectral features (photospheric and telluric absorption lines) that need to be corrected for before analyzing the optical line profiles. This step involves accurately determining the location of these absorption lines, by correctly measuring the system's radial velocity, and then using stellar templates of known systems to subtract the unwanted features. This step is crucial in the analysis because the recovered line shape and shift need to be as precise as possible for correct analysis. In addition, the the different components of the line usually overlap which introduces further uncertainty when disentangling them \citep{Banzatti2019}.

We present a novel method using machine learning to interpret the observations, which eliminates the subjectivity inherent in the current methods including Gaussian fitting. The asymmetry and overlapping components in the line shape introduce a degeneracy in the Gaussian fitting, and potentially valuable information could be lost with the Gaussian assumption. The asymmetry is an indication that emission comes from wide spread regions besides the regions where bulk emission comes from, and that part needs to be included in the models for a sound analysis of the observations. Also, the extraction of distinct components from overlapping emission could be tricky and subjective, especially if there are multiple asymmetric constituents.

One significant benefit of employing machine learning in this analysis lies in mitigating the distortion caused by background spectral features and noise. Addressing this point in detail will be postponed to a future paper, given that publicly available observations typically have these contaminants corrected for, and reintroducing them to the spectra requires meticulous handling. However, our approach promises to better quantify the relationship between uncertainties in the wind parameters and uncertainties in the background features such as the systemic velocity, stellar and atmospheric spectral templates, etc.

The fidelity of the analysis provided here is largely sensitive to the validity of the model used to describe PPD systems. The radiation sources used in the radiative transfer simulations are generic, but PPD systems in reality span a wide range of stellar, far ultraviolet, and X-ray sources depending on the type of the system. Using generic radiation sources proved useful in interpreting observations in the past \citep{Fang+2023}, and further refinement of the sources to specific systems would provide better constraints on the observed systems. We also do not include jets in our model of disk winds, and those are suspected to be responsible for one of the overlapping emission components (the HVC). Yet, we can apply the model to disks that do not display the high velocity component and those represent about half of the available sample of PPDs \citep{Fang+2018,Banzatti2019}. Nonetheless, our SBI scheme is sufficiently modular that other wind models could be used if such models became available.

We show in this paper that modern machine learning-based techniques are effective in analyzing PPD emission spectra. They offer robust analysis of complex line shapes beyond the simple Gaussian assumption. With the emergence of sophisticated models for emission spectra, SBI becomes increasingly compelling to extract parameter constraints in a computationally efficient way. We show the successful application of SBI to the analysis of simulated PPD spectra. This work shall be extended to further analyze actual observations and other complicated systems. 

\section*{Acknowledgments}
This work was supported in part by NASA grant 17-ATP17-0094 (to JG).


\begin{thebibliography}{}
\expandafter\ifx\csname natexlab\endcsname\relax\def\natexlab#1{#1}\fi
\providecommand{\url}[1]{\href{#1}{#1}}
\providecommand{\dodoi}[1]{doi:~\href{http://doi.org/#1}{\nolinkurl{#1}}}
\providecommand{\doeprint}[1]{\href{http://ascl.net/#1}{\nolinkurl{http://ascl.net/#1}}}
\providecommand{\doarXiv}[1]{\href{https://arxiv.org/abs/#1}{\nolinkurl{https://arxiv.org/abs/#1}}}

\bibitem[{{Alexander} {et~al.}(2014){Alexander}, {Pascucci}, {Andrews},
  {Armitage}, \& {Cieza}}]{Alexander+2014}
{Alexander}, R., {Pascucci}, I., {Andrews}, S., {Armitage}, P., \& {Cieza}, L.
  2014, in Protostars and Planets VI, ed. H.~{Beuther}, R.~S. {Klessen}, C.~P.
  {Dullemond}, \& T.~{Henning}, 475

\bibitem[{{Alexander}(2008)}]{Alexander2008}
{Alexander}, R.~D. 2008, \mnras, 391, L64,
  \dodoi{10.1111/j.1745-3933.2008.00556.x}

\bibitem[{{Alsing} {et~al.}(2019){Alsing}, {Charnock}, {Feeney}, \&
  {Wandelt}}]{alsing2019}
{Alsing}, J., {Charnock}, T., {Feeney}, S., \& {Wandelt}, B. 2019, \mnras, 488,
  4440, \dodoi{10.1093/mnras/stz1960}

\bibitem[{{Bai} {et~al.}(2016){Bai}, {Ye}, {Goodman}, \& {Yuan}}]{BYGY16}
{Bai}, X.-N., {Ye}, J., {Goodman}, J., \& {Yuan}, F. 2016, \apj, 818, 152,
  \dodoi{10.3847/0004-637X/818/2/152}

\bibitem[{{Ballabio} {et~al.}(2020){Ballabio}, {Alexander}, \&
  {Clarke}}]{Ballabio+2020}
{Ballabio}, G., {Alexander}, R.~D., \& {Clarke}, C.~J. 2020, \mnras, 496, 2932,
  \dodoi{10.1093/mnras/staa1767}

\bibitem[{{Bally}(2016)}]{Bally2016}
{Bally}, J. 2016, \araa, 54, 491, \dodoi{10.1146/annurev-astro-081915-023341}

\bibitem[{Banzatti {et~al.}(2019)Banzatti, Pascucci, Edwards, Fang, Gorti, \&
  Flock}]{Banzatti2019}
Banzatti, A., Pascucci, I., Edwards, S., {et~al.} 2019, \apj, 870, 76,
  \dodoi{10.3847/1538-4357/AAF1AA}

\bibitem[{Chen {et~al.}(2023)Chen, Harness, \& Melchior}]{Chen2023}
Chen, A., Harness, A., \& Melchior, P. 2023, Journal of Astronomical
  Telescopes, Instruments, and Systems, 9, 025002,
  \dodoi{10.1117/1.JATIS.9.2.025002}

\bibitem[{Dax {et~al.}(2021)Dax, Green, Gair, Macke, Buonanno, \&
  Sch{\"o}lkopf}]{dax2021}
Dax, M., Green, S.~R., Gair, J., {et~al.} 2021, Physical Review Letters, 127,
  241103, \dodoi{10.1103/PhysRevLett.127.241103}

\bibitem[{{Durkan} {et~al.}(2019){Durkan}, {Bekasov}, {Murray}, \&
  {Papamakarios}}]{Durkan2019}
{Durkan}, C., {Bekasov}, A., {Murray}, I., \& {Papamakarios}, G. 2019, arXiv
  e-prints, arXiv:1906.04032, \dodoi{10.48550/arXiv.1906.04032}

\bibitem[{{Ellerbroek} {et~al.}(2013){Ellerbroek}, {Podio}, {Kaper}, {Sana},
  {Huppenkothen}, {de Koter}, \& {Monaco}}]{Ellerbroek+2013}
{Ellerbroek}, L.~E., {Podio}, L., {Kaper}, L., {et~al.} 2013, \aap, 551, A5,
  \dodoi{10.1051/0004-6361/201220635}

\bibitem[{{Ercolano} \& {Owen}(2016)}]{Ercolano+Owen2016}
{Ercolano}, B., \& {Owen}, J.~E. 2016, \mnras, 460, 3472,
  \dodoi{10.1093/mnras/stw1179}

\bibitem[{Ercolano \& Pascucci(2017)}]{Ercolano2017}
Ercolano, B., \& Pascucci, I. 2017, {The dispersal of planet-forming discs:
  Theory confronts observations},  Royal Society Publishing,
  \dodoi{10.1098/rsos.170114}

\bibitem[{{Fang} {et~al.}(2018){Fang}, {Pascucci}, {Edwards}, {Gorti},
  {Banzatti}, {Flock}, {Hartigan}, {Herczeg}, \& {Dupree}}]{Fang+2018}
{Fang}, M., {Pascucci}, I., {Edwards}, S., {et~al.} 2018, \apj, 868, 28,
  \dodoi{10.3847/1538-4357/aae780}

\bibitem[{{Fang} {et~al.}(2023){Fang}, {Wang}, {Herczeg}, {Hashimoto}, {Xu},
  {Nemer}, {Pascucci}, {Haffert}, \& {Aoyama}}]{Fang+2023}
{Fang}, M., {Wang}, L., {Herczeg}, G.~J., {et~al.} 2023, Nature Astronomy,
  \dodoi{10.1038/s41550-023-02004-x}

\bibitem[{{Ferland} {et~al.}(2017){Ferland}, {Chatzikos}, {Guzm{\'a}n},
  {Lykins}, {van Hoof}, {Williams}, {Abel}, {Badnell}, {Keenan}, {Porter}, \&
  {Stancil}}]{Ferland+2017}
{Ferland}, G.~J., {Chatzikos}, M., {Guzm{\'a}n}, F., {et~al.} 2017, \rmxaa, 53,
  385.
\newblock \doarXiv{1705.10877}

\bibitem[{{Font} {et~al.}(2004){Font}, {McCarthy}, {Johnstone}, \&
  {Ballantyne}}]{Font+2004}
{Font}, A.~S., {McCarthy}, I.~G., {Johnstone}, D., \& {Ballantyne}, D.~R. 2004,
  \apj, 607, 890, \dodoi{10.1086/383518}

\bibitem[{{Frank} {et~al.}(2014){Frank}, {Ray}, {Cabrit}, {Hartigan}, {Arce},
  {Bacciotti}, {Bally}, {Benisty}, {Eisl{\"o}ffel}, {G{\"u}del}, {Lebedev},
  {Nisini}, \& {Raga}}]{Frank+2014}
{Frank}, A., {Ray}, T.~P., {Cabrit}, S., {et~al.} 2014, in Protostars and
  Planets VI, ed. H.~{Beuther}, R.~S. {Klessen}, C.~P. {Dullemond}, \&
  T.~{Henning}, 451, \dodoi{10.2458/azu_uapress_9780816531240-ch020}

\bibitem[{{Germain} {et~al.}(2015){Germain}, {Gregor}, {Murray}, \&
  {Larochelle}}]{Germain2015}
{Germain}, M., {Gregor}, K., {Murray}, I., \& {Larochelle}, H. 2015, arXiv
  e-prints, arXiv:1502.03509, \dodoi{10.48550/arXiv.1502.03509}

\bibitem[{{Gorti} \& {Hollenbach}(2009)}]{Gorti+2009}
{Gorti}, U., \& {Hollenbach}, D. 2009, \apj, 690, 1539,
  \dodoi{10.1088/0004-637X/690/2/1539}

\bibitem[{{G{\"u}del} {et~al.}(2014){G{\"u}del}, {Dvorak}, {Erkaev}, {Kasting},
  {Khodachenko}, {Lammer}, {Pilat-Lohinger}, {Rauer}, {Ribas}, \&
  {Wood}}]{Gudel+2014}
{G{\"u}del}, M., {Dvorak}, R., {Erkaev}, N., {et~al.} 2014, in Protostars and
  Planets VI, ed. H.~{Beuther}, R.~S. {Klessen}, C.~P. {Dullemond}, \&
  T.~{Henning}, 883, \dodoi{10.2458/azu_uapress_9780816531240-ch038}

\bibitem[{{Hahn} \& {Melchior}(2022)}]{Hahn2022_sedflow}
{Hahn}, C., \& {Melchior}, P. 2022, \apj, 938, 11,
  \dodoi{10.3847/1538-4357/ac7b84}

\bibitem[{{Hahn} {et~al.}(2022){Hahn}, {Eickenberg}, {Ho}, {Hou}, {Lemos},
  {Massara}, {Modi}, {Moradinezhad Dizgah}, {R{\'e}galdo-Saint Blancard}, \&
  {Abidi}}]{Hahn2022_simbig}
{Hahn}, C., {Eickenberg}, M., {Ho}, S., {et~al.} 2022, arXiv e-prints,
  arXiv:2211.00723, \dodoi{10.48550/arXiv.2211.00723}

\bibitem[{{Hartigan} {et~al.}(1995){Hartigan}, {Edwards}, \&
  {Ghandour}}]{Hartigan+1995}
{Hartigan}, P., {Edwards}, S., \& {Ghandour}, L. 1995, \apj, 452, 736,
  \dodoi{10.1086/176344}

\bibitem[{{Kingma} \& {Ba}(2014)}]{Adam}
{Kingma}, D.~P., \& {Ba}, J. 2014, arXiv e-prints, arXiv:1412.6980,
  \dodoi{10.48550/arXiv.1412.6980}

\bibitem[{{Kobyzev} {et~al.}(2019){Kobyzev}, {Prince}, \&
  {Brubaker}}]{Kobyzev2019}
{Kobyzev}, I., {Prince}, S. J.~D., \& {Brubaker}, M.~A. 2019, arXiv e-prints,
  arXiv:1908.09257, \dodoi{10.48550/arXiv.1908.09257}

\bibitem[{{Lemos} {et~al.}(2023){Lemos}, {Parker}, {Hahn}, {Ho}, {Eickenberg},
  {Hou}, {Massara}, {Modi}, {Moradinezhad Dizgah}, {Regaldo-Saint Blancard}, \&
  {Spergel}}]{lemos2023}
{Lemos}, P., {Parker}, L., {Hahn}, C., {et~al.} 2023, arXiv e-prints,
  arXiv:2310.15256.
\newblock \doarXiv{2310.15256}

\bibitem[{{Liang} {et~al.}(2023{\natexlab{a}}){Liang}, {Melchior}, {Hahn},
  {Shen}, {Goulding}, \& {Ward}}]{Liang2023b}
{Liang}, Y., {Melchior}, P., {Hahn}, C., {et~al.} 2023{\natexlab{a}}, arXiv
  e-prints, arXiv:2307.07664, \dodoi{10.48550/arXiv.2307.07664}

\bibitem[{{Liang} {et~al.}(2023{\natexlab{b}}){Liang}, {Melchior}, {Lu},
  {Goulding}, \& {Ward}}]{Liang2023a}
{Liang}, Y., {Melchior}, P., {Lu}, S., {Goulding}, A., \& {Ward}, C.
  2023{\natexlab{b}}, \aj, 166, 75, \dodoi{10.3847/1538-3881/ace100}

\bibitem[{{Manara} {et~al.}(2023){Manara}, {Ansdell}, {Rosotti}, {Hughes},
  {Armitage}, {Lodato}, \& {Williams}}]{Manara+2023}
{Manara}, C.~F., {Ansdell}, M., {Rosotti}, G.~P., {et~al.} 2023, in
  Astronomical Society of the Pacific Conference Series, Vol. 534, Astronomical
  Society of the Pacific Conference Series, ed. S.~{Inutsuka}, Y.~{Aikawa},
  T.~{Muto}, K.~{Tomida}, \& M.~{Tamura}, 539,
  \dodoi{10.48550/arXiv.2203.09930}

\bibitem[{{Melchior} {et~al.}(2023){Melchior}, {Liang}, {Hahn}, \&
  {Goulding}}]{Melchoir+2023}
{Melchior}, P., {Liang}, Y., {Hahn}, C., \& {Goulding}, A. 2023, \aj, 166, 74,
  \dodoi{10.3847/1538-3881/ace0ff}

\bibitem[{{Natta} {et~al.}(2014){Natta}, {Testi}, {Alcal{\'a}}, {Rigliaco},
  {Covino}, {Stelzer}, \& {D'Elia}}]{Natta+2014}
{Natta}, A., {Testi}, L., {Alcal{\'a}}, J.~M., {et~al.} 2014, \aap, 569, A5,
  \dodoi{10.1051/0004-6361/201424136}

\bibitem[{{Nemer} \& {Goodman}(2024)}]{Nemer+2023}
{Nemer}, A., \& {Goodman}, J. 2024, \apj, 961, 122,
  \dodoi{10.3847/1538-4357/ad0a8d}

\bibitem[{{Papamakarios} {et~al.}(2017){Papamakarios}, {Pavlakou}, \&
  {Murray}}]{Papamakarios2017}
{Papamakarios}, G., {Pavlakou}, T., \& {Murray}, I. 2017, arXiv e-prints,
  arXiv:1705.07057, \dodoi{10.48550/arXiv.1705.07057}

\bibitem[{{Pascucci} {et~al.}(2023){Pascucci}, {Cabrit}, {Edwards}, {Gorti},
  {Gressel}, \& {Suzuki}}]{Pascucci+2023}
{Pascucci}, I., {Cabrit}, S., {Edwards}, S., {et~al.} 2023, in Astronomical
  Society of the Pacific Conference Series, Vol. 534, Astronomical Society of
  the Pacific Conference Series, ed. S.~{Inutsuka}, Y.~{Aikawa}, T.~{Muto},
  K.~{Tomida}, \& M.~{Tamura}, 567, \dodoi{10.48550/arXiv.2203.10068}

\bibitem[{{Pascucci} \& {Sterzik}(2009)}]{Pascucci+Sterzik2009}
{Pascucci}, I., \& {Sterzik}, M. 2009, \apj, 702, 724,
  \dodoi{10.1088/0004-637X/702/1/724}

\bibitem[{{Rigliaco} {et~al.}(2013){Rigliaco}, {Pascucci}, {Gorti}, {Edwards},
  \& {Hollenbach}}]{Rigliaco+2013}
{Rigliaco}, E., {Pascucci}, I., {Gorti}, U., {Edwards}, S., \& {Hollenbach}, D.
  2013, \apj, 772, 60, \dodoi{10.1088/0004-637X/772/1/60}

\bibitem[{{Simon} {et~al.}(2016){Simon}, {Pascucci}, {Edwards}, {Feng},
  {Gorti}, {Hollenbach}, {Rigliaco}, \& {Keane}}]{Simon+2016}
{Simon}, M.~N., {Pascucci}, I., {Edwards}, S., {et~al.} 2016, \apj, 831, 169,
  \dodoi{10.3847/0004-637X/831/2/169}

\bibitem[{{Smith} \& {Topin}(2017)}]{Smith2017}
{Smith}, L.~N., \& {Topin}, N. 2017, arXiv e-prints, arXiv:1708.07120,
  \dodoi{10.48550/arXiv.1708.07120}

\bibitem[{Tabak \& Turner(2013)}]{Tabak2013}
Tabak, E., \& Turner, C. 2013, Communications on Pure and Applied Mathematics,
  66, 145, \dodoi{10.1002/cpa.21423}

\bibitem[{Tabak \& Vanden-Eijnden(2010)}]{Tabak2010}
Tabak, E., \& Vanden-Eijnden, E. 2010, Communications in Mathematical Sciences,
  8, 217, \dodoi{10.4310/CMS.2010.v8.n1.a11}

\bibitem[{{Talts} {et~al.}(2018){Talts}, {Betancourt}, {Simpson}, {Vehtari}, \&
  {Gelman}}]{Talts2018}
{Talts}, S., {Betancourt}, M., {Simpson}, D., {Vehtari}, A., \& {Gelman}, A.
  2018, arXiv e-prints, arXiv:1804.06788, \dodoi{10.48550/arXiv.1804.06788}

\bibitem[{Tejero-Cantero {et~al.}(2020)Tejero-Cantero, Boelts, Deistler,
  Lueckmann, Durkan, Gonçalves, Greenberg, \& Macke}]{tejero-cantero2020sbi}
Tejero-Cantero, A., Boelts, J., Deistler, M., {et~al.} 2020, Journal of Open
  Source Software, 5, 2505, \dodoi{10.21105/joss.02505}

\bibitem[{{Weber} {et~al.}(2020){Weber}, {Ercolano}, {Picogna}, {Hartmann}, \&
  {Rodenkirch}}]{Weber+2020}
{Weber}, M.~L., {Ercolano}, B., {Picogna}, G., {Hartmann}, L., \& {Rodenkirch},
  P.~J. 2020, \mnras, 496, 223, \dodoi{10.1093/mnras/staa1549}

\bibitem[{Wong {et~al.}(2020)Wong, Contardo, \& Ho}]{wong2020}
Wong, K. W.~K., Contardo, G., \& Ho, S. 2020, Physical Review D, 101, 123005,
  \dodoi{10.1103/PhysRevD.101.123005}

\bibitem[{{Zhang} {et~al.}(2021){Zhang}, {Bloom}, {Gaudi}, {Lanusse}, {Lam}, \&
  {Lu}}]{zhang2021}
{Zhang}, K., {Bloom}, J.~S., {Gaudi}, B.~S., {et~al.} 2021, \aj, 161, 262,
  \dodoi{10.3847/1538-3881/abf42e}

\end{thebibliography}

\end{CJK*}
\end{document}